\documentclass[10pt]{iopart}
\usepackage{iopams}

\expandafter\let\csname equation*\endcsname\relax
\expandafter\let\csname endequation*\endcsname\relax

\usepackage{amssymb}
\usepackage{amsmath}
\usepackage{soul}
\usepackage{cite}
\usepackage{enumerate}
\usepackage{empheq}
\usepackage{fancybox}
\usepackage{etoolbox}
\usepackage{hyperref}
\usepackage{color}
\makeatletter
\def\@mkboth#1#2{}
\newlength\appendixwidth
\preto\appendix{\addtocontents{toc}{\protect\patchl@section}}
\newcommand{\patchl@section}{%
  \settowidth{\appendixwidth}{\textbf{Appendix }}%
  \addtolength{\appendixwidth}{1.5em}%
  \patchcmd{\l@section}{1.5em}{\appendixwidth}{}{\ddt}%
}
\makeatother

\begin{document}

\title[Crowded Random Manhattan Lattice]{Tracer Diffusion on a Crowded Random Manhattan Lattice}

\author{Carlos Mej\'{i}a-Monasterio$^{1}$, Sergei Nechaev$^{2,3}$, Gleb Oshanin$^{2,4}$ \& Oleg Vasilyev$^{5,6}$\footnote{Corresponding author}}
\address{$^1$ School of Agricultural, Food and Biosystems Engineering, Technical University of Madrid, Av. Complutense s/n, 28040 Madrid, Spain}
\address{$^2$
Interdisciplinary Scientific Center
J.-V. Poncelet (UMI CNRS 2615), Bolshoy Vlasyevskiy Lane 11, 119002 Moscow, Russia}
\address{$^3$ Lebedev Physical Institute RAS, 119991, Moscow, Russia}
\address{$^4$ Sorbonne Universit\'e, CNRS, Laboratoire de Physique 
Th\'eorique de la Mati\`ere Condens\'ee (UMR CNRS 7600),
 4 Place Jussieu, 75252 Paris Cedex 05, France}
 \address{$^5$ Max-Planck-Institut f\"ur Intelligente Systeme, 
 Heisenbergstr. 3, D-70569, Stuttgart, Germany} 
\address{$^6$ IV. Institut f\"ur Theoretische Physik, 
Universit\"at Stuttgart, Pfaffenwaldring 57, D-70569 Stuttgart, Germany}

\begin{abstract}
We study by extensive numerical simulations 
the dynamics of a hard-core tracer particle (TP) in presence of two competing types of disorder - \textit{frozen} convection flows on a square random Manhattan lattice and a crowded \textit{dynamical} environment formed by a lattice gas of mobile hard-core particles. The latter
perform lattice random walks, 
constrained 
by a single-occupancy condition of each lattice site, and are either insensitive to random flows (model A) or
choose the jump directions as dictated by the local directionality of bonds of the random Manhattan lattice (model B). 
We focus on the TP  disorder-averaged mean-squared displacement, (which shows a super-diffusive behaviour $\sim t^{4/3}$, $t$ being time, in all the cases studied here), on higher 
moments of the TP displacement, and on the probability distribution of the TP position $X$ along the $x$-axis. 
Our analysis evidences that
in absence of the lattice gas particles the latter has a Gaussian central part $\sim \exp(- u^2)$,  where $u = X/t^{2/3}$, and exhibits
slower-than-Gaussian tails $\sim \exp(-|u|^{4/3})$  for sufficiently large $t$ and $u$. 
Numerical data convincingly demonstrate that in presence of a crowded environment 
the central Gaussian part and non-Gaussian tails of the distribution persist for both models.
 \end{abstract}

\eads{carlos.mejia@upm.es, nechaev@lptms.upsud.fr,  oshanin@lptmc.jussieu.fr, vasilyev@fluids.mpi-stuttgart.mpg.de}
\vspace{10pt}


\pagebreak

\section{Introduction}

In many realistic systems encountered across several disciplines -  e.g.,  physics, chemistry, molecular and cellular biology, - 
random motion of tracer particles 
takes place in presence of disorder, either temporal or spatial, which may originate from a variety of different factors 
\cite{bern,kl,hav,comtet,bouch,gleb,ben,stas,yuval,felix,igor,ralf,oli}. 
Understanding the impact of disorder on dynamics is thus a challenging issue, which has important conceptual and practical implications.  

 Quenched (frozen) spatial disorder which 
 entails a temporal trapping of a tracer particle (TP) at some  positions,
 often produces an anomalous sub-diffusive behaviour, especially in low-dimensional systems. Here, the TP trajectories are spatially more confined than the trajectories of a standard Brownian motion. As a consequence, the disorder-averaged mean-squared displacement (DA MSD) 
  behaves as $<\overline{X^2(t)}> \sim t^{\gamma}$, with $t$ being time and $\gamma$ - the dynamical exponent which  is less than unity. Here and henceforth, the bar denotes averaging over thermal histories while the angle brackets stand for averaging over disorder.
  Striking examples of such a dynamical behaviour are provided by, e.g.,  the so-called  Sinai diffusion in one-dimensional systems \cite{sinai} (see also Refs. \cite{hav,comtet,bouch,gleb}) in which the DA MSD grows as $<\overline{X^2(t)}> \sim \ln^4 t$ (i.e., formally, $\gamma = 0$), Sinai diffusion in presence of a constant external bias \cite{kozlov,derrida} or migration of excited states along a one-dimensional array of randomly placed donor centres \cite{bern,gleb}. 
  In this latter example the dynamical exponent $\gamma$ is non-universal and equals the mean density of donor centres times the characteristic length-scale of the distance-dependent (exponential) transfer rate. If this product is less than unity, a sub-diffusive motion takes place. Two other examples concern diffusion in the "impurity
band" \cite{anderson}
  and the so-called Random Trap model \cite{bouch2,cecile,eli}. Here, as well,  $\gamma$ is non-universal and is less than unity in some region of the parameter space.  In higher-dimensional systems, diffusion in presence of such a disorder typically becomes normal (see, however, Refs. \cite{anderson,fisher,b1}) and the disorder affects only the value of the diffusion coefficient. Diffusion is also normal in the asymptotic large-$t$ limit in one-dimensional systems with a periodic disorder. Here, however, the value of the 
  diffusion coefficient may exhibit strong sample-to-sample fluctuations and thus 
 have non-trivial statistical properties, such that  the averaged diffusion coefficient 
  will not be representative of the actual behaviour (see, e.g., Ref. \cite{david}).  The large-$t$ relaxation of the diffusion coefficient to its asymptotic value may shed some light on the kind of disorder one is dealing with \cite{dean}.

\begin{figure}[htb!]
\begin{center}
\includegraphics[width=1.0\hsize]{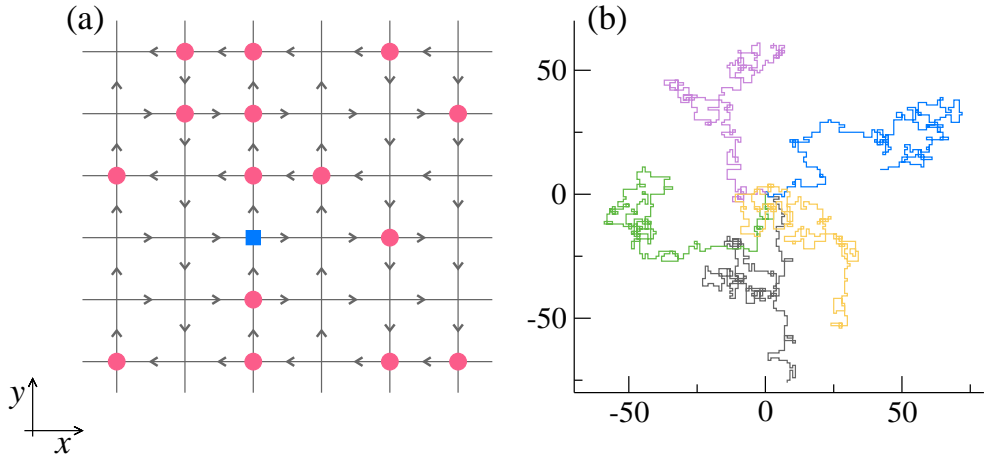}
\end{center}
\caption{{\em Random Manhattan lattice and the TP trajectories}. Panel (a). A realisation of a random Manhattan lattice - a square lattice decorated in a random fashion with arrows,  indicating 
the possible jump directions. Jumps against an arrow are not permitted in our model. The directionality of each arrow 
is fixed along each street (East-West) and an avenue (North-South) along their entire, infinite in both directions length,
and fluctuates randomly from a street (an avenue) to a street (an avenue). The pattern of arrows is frozen and does not vary with time.
 A square (blue) indicates the TP instantaneous position, while the circles (red) denote the instantaneous positions of the LG particles. Panel (b). Five individual TP trajectories on a random Manhattan lattice in absence of the LG particles.}
\label{fig1}
\end{figure}

Random frozen convection (velocity) flows most often produce a  
super-diffusion with $\gamma > 1$. To name just two such situations, we mention a model in which a TP
is passively advected by quenched,
layered, randomly-oriented flows  (say, along the $x$-axis) and undergoes a normal diffusion in the direction perpendicular to them (i.e., along the $y$-axis), 
as well as its generalisation - a random Manhattan lattice (see Fig. \ref{fig1}), in which the orientation of convection flows randomly fluctuates both along the streets and avenues (i.e., along both $x$- and $y$-axes). The former model
was introduced originally 
for the analysis of conductivity of inhomogeneous media in a strong magnetic field \cite{dykhne} and of the dynamics
of solute in a stratified porous medium with flow parallel to the bedding \cite{MdM}. In such a setting, usually referred to as the Matheron - de Marsily (MdM) model according to the names of authors of Ref. \cite{MdM}, 
the TP dynamics  in the flow direction (along the $x$-axis) is characterised by a super-diffusive law of the form $<\overline{X^2(t)}> \sim t^{3/2}$, i.e., $\gamma = 3/2$. Many interesting generalisations and more details on the available analytical and numerical results can be found 
in Refs. \cite{redner,ledoussal2,ledoussal3,crisanti,blumen3,ledoussal4,jespersen,majumdar,alessio}.  Diffusion of a single TP on a square random Manhattan lattice has been analysed in Refs. \cite{redner,ledoussal2}. It was shown, by using simple analytical arguments and a numerical analysis, that in this case the DA MSD also exhibits a super-diffusive behaviour, but with a somewhat smaller dynamical exponent $\gamma = 4/3$, i.e., the DA MSD of the $x$-component
of the TP position obeys
$<\overline{X^2(t)}> \sim t^{4/3}$.   This model has been also widely studied in different contexts in mathematical literature (see, e.g., Ref. \cite{ledger}).  A generalisation of a random Manhattan lattice was invoked as an example of a plausible geometric disorder
 in a recent analysis 
of the localisation length exponent for plateau transition in quantum Hall effect \cite{ser}. This latter setting, however, is clearly more complicated than the MdM model with the layered flows and the theoretical progress here is rather limited;  the behaviour beyond the temporal evolution of a DA MSD is still largely unknown. 

Dynamical disorder emerges naturally when the TP's transition rates fluctuate
 randomly in time, as it happens, for instance, 
 in physical processes underlying the so-called diffusing-diffusivity models 
 \cite{chub,chech,greb,fulvio,yael,we,eli2} or the dynamic percolation \cite{dp1,dp2,dp3}.
 Another pertinent case concerns the situations when the TP 
 evolves in a dynamical environment of
mobile steric obstacles - interacting crowders which impede its dynamics (see, e.g., Refs. \cite{felix,igor,ralf}). 
 A paradigmatic example of such a situation is provided by a TP diffusion
in lattice gases of hard-core particles, which undergo the so-called simple exclusion process (see Ref. \cite{oli} for a recent review), i.e., perform lattice random walks subject to the constraint that each lattice site can be at most singly occupied. It is well-known that in 
such an environment  the particles' dynamics is strongly correlated. 
These correlations are especially
 important and cause an essential departure from standard diffusive motion
  in two cases: a) in one-dimensional geometry
   - the so-called single-files,   in which the particles cannot bypass each other and the initial order of particles is preserved at all times; and b) on ramified comb-like structures consisting of an infinitely long single-file backbone with infinitely long single-file side branches, which permit for some re-ordering of particles.
   In single-files, the TP mean-squared displacement exhibits an anomalous sub-diffusive behaviour
 $\overline{X^2(t)} \sim t^{1/2}$. This striking result was first obtained analytically by Harris  \cite{harris} (see Refs. \cite{taloni,michael} for a review), and holds also for all the cumulants of $X(t)$ \cite{poncet1,kirone} and  
in case of multiple TPs \cite{poncet2,ooshida,poncet3}. On crowded comb-like structures, the TP mean-squared displacement exhibits a variety of sub-diffusive transients and, in some cases, an ultimate sub-diffusive behaviour \cite{combs}. 
 On higher-dimensional lattices, the TP dynamics becomes diffusive in the large-$t$ limit with  the effective diffusion coefficient being a non-trivial function of the density of crowders and other pertinent parameters \cite{kit,koi,kehr,tahir,bei,oliv1,oliv2}. 
 This non-trivial behaviour of the diffusion coefficient is associated with the enhanced probability of backward jumps - in a crowded environment, for any particle it is more probable to return back to the site it just left vacant, than to keep on going farther away
  \cite{kit,koi,kehr,tahir,bei,oliv1,oliv2}.
  
 Meanwhile, a considerable knowledge is accumulated 
 through case-by-case  theoretical and numerical analyses  
 of the TP
dynamics in a variety of model systems with either quenched or dynamical
disorder (see, e.g.,  Refs.\cite{bern,kl,hav,comtet,bouch,gleb,ben,stas,yuval,felix,igor,ralf,oli} and references therein). 
On contrary, still little is known 
about the TP diffusion in situations in which several types of disorder are acting simultaneously. 
To the best of our knowledge, the only work 
addressing specifically  this question is recent Ref. \cite{michael}, which  focused on the TP random motion 
in single-files of hard-core particles having 
a broad \textit{scale-free}  
distribution of waiting times, e.g., due to a temporal trapping of particles. Using some subordination arguments and numerical analysis, it was shown that here a combined effect of the disorder in transition rates
and of the dynamical environment leads to a severe slowing-down of the TP random motion. Namely, the DA MSD of the TP follows $\langle \overline{X^2(t)}\rangle \sim \ln^{1/2}(t)$, i.e., exhibits an essentially slower growth with time than the one taking place
in systems in which either type of disorder is present alone. In case when a characteristic mean waiting time exists, i.e., the distribution is not scale-free, but the second moment diverges, the DA MSD grows faster than logarithmically, $\langle \overline{X^2(t)} \rangle \sim t^{\gamma}$ with $\gamma < 1/2$, but still slower than the above mentioned Harris' law.

This paper is devoted to a question of the TP dynamics in presence of  two interspersed types of disorder,  
which act concurrently and compete with each other. 
We consider the TP random motion 
 subject to \textit{quenched} random convection flows, 
which prompt a super-diffusive behaviour of the TP, in a \textit{dynamical} environment which 
is damping its random motion.  
More specifically, we study here by extensive numerical simulations 
the dynamics of a TP which evolves 
on a square random Manhattan lattice of frozen (i.e. not varying in time) convection flows in presence of a lattice gas (LG) of mobile hard-core particles.
The latter 
are either insensitive to convection flows, performing standard   
random walks among the nearest-neighbouring sites of a lattice with the probability $1/4$ to go in any direction (Model A), or follow the convection flows (similarly to the TP) by
choosing randomly between the two directions prescribed by a local directionality of bonds of the random Manhattan lattice (Model B). In the latter case the backward jumps of any
LG particle are completely suppressed. The backward jumps of the TP are forbidden in both models.  
For both models, the TP and the LG particles obey a
simple exclusion constraint, which effectively correlates the TP random motion and the evolution of LG particles. 
We focus on such characteristics of the TP dynamics as its DA MSD, and generally, the moments of arbitrary order, the distribution of its position at time moment $t$ averaged over disorder, as well as the time evolution of the kurtosis of this distribution. We also address a question of the sample-to-sample fluctuations and analyse the MSD of the TP and the probability distribution of its position for several fixed realisations of disorder.

The paper is outlined as follows:  In Sec. \ref{model} we define the model under study and introduce basic notations. In Sec. \ref{TP} we discuss dynamics of a single TP in absence of the LG particles, appropriately revisiting the arguments presented in Refs.\cite{redner,ledoussal2}.  
We also 
present here
 results of numerical simulations for the DA MSD and for higher moments of the TP displacement, as well for the disorder-averaged probability distribution of  the TP position along the $x$-axis. This sets an instructive framework for the analysis of the TP dynamics in presence of LG particle. We close Sec. \ref{TP} addressing the issue of sample-to-sample fluctuations and also 
examine the spectral properties of the TP trajectories, which reveal several interesting features.
In Sec. \ref{particles} we consider the TP dynamics in presence of LG particles for both Model A and Model B. 
 Finally, in Sec. \ref{conc} we conclude with a brief recapitulation of our results.
 
\section{Model}
\label{model}

Consider a two-dimensional random Manhattan lattice (see Fig. \ref{fig1}), i.e., 
an infinite in both directions square lattice with unit spacing, decorated with arrows in such a way that directionality of each of them
is fixed along each street (East-West) and an avenue (North-South) for their entire  length, but whose orientation  varies randomly from a street (an avenue) to a street (an avenue). 

Let an integer $n$, $n \in (-\infty,\infty)$, numerate the columns (avenues) of the lattice, and an integer $m$, $m \in (-\infty,\infty)$, - the rows (streets), respectively. Then,
the pattern of arrows in a given frozen realisation of convection flows is specified by 
assigning to each lattice site (with integer coordinates $(n,m)$) a pair of quenched random, mutually uncorrelated 
"bias" variables $\eta_n$ and $\zeta_m$. We use a convention that $\eta_n = + 1$ if an arrow points to the North, and $\eta_n = -1$, otherwise; and $\zeta_m = + 1$ if an arrow points to the East, and $\zeta_m = -1$, otherwise. 
We focus solely 
on the case when there is no \textit{global} bias; that being, $\eta_n$ and $\zeta_m$ assume the values $\pm 1$ with equal probabilities, which implies that $\langle \eta_n \rangle = \langle \zeta_m \rangle = 0$. Furthermore, we stipulate that there are no correlations between the directions of arrows at $n$ and $n'$, and at $m$ and $m'$, i.e., 
\begin{align}
\label{corr}
\langle \eta_n  \eta_{n'} \rangle &= \delta_{n,n'}, \nonumber\\
\langle \zeta_m  \zeta_{m'} \rangle &= \delta_{m,m'}, 
\end{align}
where $\delta_{a,b}$ is the Kronecker-delta, such that $\delta_{a,b} =  1$ for $a=b$, and equals zero otherwise. 
\subsection{A single tracer particle}

At time moment $t = 0$ ($t$ is a discrete time variable, $t=0,1,2, \ldots$), we introduce the TP at the origin of the lattice and let it move, at each tick of the clock, according to the following rules:\\
-- at each discrete time instant $t$, we toss a two-sided "coin" $\xi_{t}$ which can assume, (with equal probabilities $=1/2$), the values $+1$ and $-1$.\\
-- being at position ${\vec R}_{t} = (X_{t},Y_{t})$, (where $X_t$ and $Y_t$ are the projections of ${\vec R}_{t}$ on the $x$- and $y$-axes), 
the TP is moved, after choosing the value of $\xi_{t}$, to a new position
\begin{align}
\label{R}
{\vec R}_{t+1} = {\vec R}_{t}  + {\vec \delta_{t}} \,,
\end{align}
where the vectorial  
increment ${\vec \delta_{t}}$  is defined as
\begin{align}
{\vec \delta_{t}} = \frac{\left(1+\xi_{t}\right)}{2} \, \zeta_{Y_{t}}\, {\vec e}_x + \frac{\left(1- \xi_{t}\right)}{2} \, \eta_{X_{t}}\, {\vec e}_y  \,,
\end{align}
with $ {\vec e}_x$ and ${\vec e}_x$ being the unit vectors in the $x$- and $y$-directions, respectively. The expression  \eqref{R} can also be conveniently rewritten
 in form of two coupled, non-linear recursion relations for the integer-valued components $X_t$ and $Y_t$ :
\begin{align}
\label{2}
X_{t+1} = X_{t} +  \frac{\left(1+\xi_{t}\right)}{2} \, \zeta_{Y_{t}} \,, \,\,\,
Y_{t+1} = Y_{t} +  \frac{\left(1-\xi_{t}\right)}{2} \, \eta_{X_{t}} \,.
\end{align}
Therefore, once (with probability $1/2$) $\xi_{t}=1$, the TP is moved onto the neighbouring site along the $x$-axis in the 
direction prescribed by $\zeta_{Y_t}$, and does not change its position along the $y$-axis. 
Conversely, if $\xi_{t}=-1$, 
the TP is moved on a unit distance along the $y$-axis in the direction prescribed by  $\eta_{X_{t}}$, and does not change its position along the $x$-axis. 
We recall that the ensuing motion of the TP as defined by the recursion relations \eqref{2}
is super-diffusive, with the dynamical exponent $\gamma = 4/3$ \cite{redner,ledoussal2}.

We note parenthetically that it may apparently be possible to find an equivalent two-dimensional 
model in the continuum space and time limit, write down coupled Langevin equations for the time evolution of the components and, eventually, define the associated Fokker-Planck equation obeyed by the probability ${\Pi}_t(X,Y)$ of finding the TP at position $(X,Y)$ at time moment $t$ for a given realisation of disorder. We will address 
this question in our following work.
Second, it was claimed
 in Refs. \cite{redner,ledoussal2} that at a coarse-grained level the TP dynamics on a random Manhattan lattice becomes equivalent to a Brownian motion in continuum, in  a  divergenceless  random velocity field with power-law decay of the velocity correlation function. We however remark that going to a continuum limit necessitates a generalisation of the model studied here; in our settings, the jumps against an arrow are not permitted 
 which tacitly presumes that
 the force acting on the particle along  a given bond is infinitely large. 
 Therefore, one has to allow for the jumps against an arrow and let them occur with a smaller (but finite) 
 probability, than the probability of the jumps along an arrow. This is tantamount to considering finite forces. We, however, do not expect any substantial change in the dynamics in the finite force case,  as compared to our model. 

The algorithm of our numerical simulations of the TP dynamics on a random Manhattan lattice follows the relations \eqref{2}. We generate trajectories along the $x$- abd $y$-axes of a given length $t$, for a given set of thermal variables $\{\xi_t\}$ and a given realisation of "bias" variables $\eta_n$ and $\zeta_m$.
The obtained individual trajectories are stored and the characteristic properties of interest - the moments of the TP displacement and the distribution function of the TP position - are evaluated by 
averaging over different realisations of trajectories.
 Averaging is first 
performed over $10^4$  trajectories generated for a fixed realisation of a random Manhattan lattice, and then the procedure is repeated for $2 \times 10^5$ realisations of disorder. 
Simulations are performed for lattices containing $L \times L$ sites with $L = 2 \times 10^6$. Care is taken that neither of the TP trajectories reaches the boundaries of the lattice within the observation time, such that the finite-size effects do not matter.  For the lattice size used in our numerical modelling, this permits us to safely explore the TP dynamics for times up to $t = 10^6$. Lastly, we also analyse the sample-to-sample fluctuations and, in particular, address a question of the TP dynamics in presence of a single fixed realisation of disorder. In this case, for a given random realisation of disorder we run $2 \times 10^7$ trajectories. 

\subsection{The TP dynamics on a crowded random Manhattan lattice}

The TP dynamics on a random Manhattan lattice populated with $N - 1$ lattice gas particles is analysed numerically. 
Due to a significant number of the particles involved, we are only able to consider square lattices 
  with the maximal linear extent $L = 2 \times 10^3$. 
  This means that the maximal time $t$, 
  until which the finite-size effects can be discarded, 
  is of order of $4 \times 10^3$. 
  Moreover, due to
  computational limitations,  we record only $50$ TP trajectories for each given realisation of disorder, and average over 
 $10^3$ realisations of disorder. Such a statistical sample appears to be sufficiently large to probe the behaviour of
 the  DA MSD of the TP,  but does not permit us to make absolutely conclusive statements about the shape of the distribution function. Nonetheless, our numerical data rather convincingly demonstrate that the overall behaviour of the latter is very similar to the one observed for the TP dynamics in absence of the LG particles, in which case a more ample statistical analysis has been performed.

 The simulations are performed as follows:
We first place the TP at the origin of a lattice and then 
distribute $N-1$ hard-core particles among the remaining sites   by 
placing a LG particle at each lattice site, at random, with probability $\rho =N/L^2$. The latter parameter defines the mean density of particles in the system;   in our simulations,  we study the TP dynamics for
  nine values of $\rho$, $\rho = 0.1, 0.2, 0.3, \ldots, 0.9$.   

After the particles are introduced into the system, they are let to move randomly subject to a single-occupancy constraint. We distinguish between two possible scenarios:

\subsubsection{Model A.}

In model $A$ we suppose that all the LG particles 
are not \textit{sensitive} to the frozen pattern of convection flows and perform 
symmetric random walks, subject to the constraint that there may be at most a single particle (i.e., either the TP or a LG particle) at each lattice site. On contrary, for the TP the choice of the jump direction is dictated by the arrows present at the site it occupies at time moment $t$. As described above, the TP chooses at random between the two arrows outgoing from the site it occupies.
In this case, the TP  (which exhibits a super-diffusive motion in absence of the LG particles) is not identical to the LG particles and 
moves in a quiescent "fluid" of hard-core particles which exerts some frictional force on it. Note that here the backward jumps are forbidden for the TP only.

More specifically, at each step we select at random a particle, which can be either a TP or a LG particle, and let it choose the jump direction: if the selected particle is a TP,  it chooses at random between the two arrows. Conversely, 
a LG particle chooses at random one among four neighbouring sites with probability $=1/4$. The jump of a TP or a LG particle is fulfilled, once the target site is empty at this time instant; 
otherwise, the particle remains at its position.  
The time $t$ is increased by unity after repeating such a procedure $N$ times, such that
all $N$ particles present in the system, on average, have a chance 
to change their positions. 

We have already mentioned  that in this model  the dynamics of LG particles is rather non-trivial due to 
an enhanced probability of backward jumps; it means that a particle which jumps onto an empty target site will most likely return on the next time step to the site it just left vacant, then will keep on going away from it. Even in absence of  the TP and random convection flows acting on it, 
this circumstance results in a non-trivial dependence of the self-diffusion coefficient $D_{tp}$ of any tagged particle on the overall density of the LG particles. This dependence  is known only in an approximate form (see, e.g., Refs. \cite{oli} and \cite{kit,koi,kehr,tahir,bei,oliv1,oliv2}). 
The available exact results concern the leading, in the dense limit $\rho \simeq 1$, behaviour of the self-diffusion coefficient $D_{tp} \simeq (1-\rho)/(4 (\pi - 1))$ \cite{brum} and of the mobility $\mu_{tp} \simeq \beta (1 - \rho)/(4 (\pi - 1))$ \cite{oli7} of a tagged particle subject to a vanishingly small external force, with 
$\beta$ being the reciprocal temperature. The appearance of the Archimedes' irrational number "$\pi$" seems  astonishing and points on a non-trivial behaviour.

\subsubsection{Model B.}
In this model, we suppose that all the particles in the system are identical. It means that both the TP and the LG particles move on the lattice subject to a single-occupancy constraint and obey the rules of the random Manhattan lattice,  by following the jump directions prescribed by the arrows. 

Note that in this model the backward jump probability is equal to zero for all the particles, both for the LG particles and the TP. 
As a consequence, we expect that here the environment in which the TP moves is a kind of a "turbulent" fluid, in which all the particles exhibit a super-diffusive motion. Hence, we may expect that 
the environment becomes perfectly stirred at sufficiently large times, such that the time $t$ gets simply rescaled by the frequency  $(1 - \rho)$ of successful jump events,  
 (which is not the case for Model A). 
We are going to verify if this is the case in what follows.

\section{Dynamics of a single tracer particle}
\label{TP}

\subsection{Disorder-averaged mean-squared displacement}

In order to calculate the DA MSD of a single TP moving on a random Manhattan lattice in absence of the LG particles,
we suitably revisit the arguments presented in Refs.\cite{redner,ledoussal2}. The latter were based on an estimate of  typical fluctuations of sums of quenched random variables $\eta_n$ and $\zeta_m$, and a plausible closure relation. Here, we pursue a bit different line of thought.

First, we  "solve" the recursions in eqs. \eqref{2} to get, for $t \geq 1$,
\begin{align}
\label{7}
X_{t} &= \sum_{\tau=0}^{t-1} \frac{\left(1+\xi_{\tau}\right)}{2} \, \zeta_{Y_{\tau}} \,, \,\,\,
Y_{t} = \sum_{\tau=0}^{t-1} \frac{\left(1-\xi_{\tau}\right)}{2} \, \eta_{X_{\tau}} \,,
\end{align}
with the initial condition $X_0 = Y_0 = 0$.
Expressions \eqref{7} define the TP positions $X_t$ and $Y_t$ for any $t$, for fixed realisations of thermal noises $\xi_{t}$ and "biases" $\eta_{n}$ and $\zeta_{m}$.

We concentrate on the $x$-component and write down formally its squared value:
\begin{align}
\label{x}
X_{t}^2 = \sum_{\tau=0}^{t-1} \frac{\left(1+\xi_{\tau}\right)^2}{4} \, \zeta^2_{Y_{\tau}} + 2 \sum_{\tau=0}^{t-2} \sum_{\tau' = \tau+1}^{t-1}  \frac{\left(1+\xi_{\tau}\right)}{2} \frac{\left(1+\xi_{\tau'}\right)}{2}  \zeta_{Y_{\tau}}  \zeta_{Y_{\tau'}} \,.
\end{align}
Let the bar denote averaging over $\xi_{\tau}$-s, which amounts to averaging over thermal histories,
and the angle brackets - averaging over random variables $\eta_n$ and $\zeta_m$, i.e., averaging over quenched disorder. 
Consider the averaged first sum in the right-hand-side (rhs) of eq. \eqref{x}. Noticing that
$ \zeta^2_{Y_{\tau}} \equiv 1$, i.e., is not fluctuating, we realise that the averaged first sum is simply
\begin{align}
\left \langle \overline{ \sum_{\tau=0}^{t-1} \frac{\left(1+\xi_{\tau}\right)^2}{4} \, \zeta^2_{Y_{\tau}}} \right \rangle = \sum_{\tau=0}^{t-1} \overline{  \frac{\left(1+\xi_{\tau}\right)^2}{4} } = \frac{t}{2} \,.
\end{align}
Hence,  the contribution of the averaged first sum to the DA MSD of the TP 
along the $x$-axis is that of 
a standard, discrete-time random walk (with the diffusion coefficient $D= 1/4$)
on a two-dimensional undecorated square lattice.

Focus on the summand in the second term in the rhs of eq. \eqref{x} and write down formally its averaged value:
\begin{align}
\label{mu}
2 \left \langle \overline{   \frac{\left(1+\xi_{\tau}\right)}{2} \frac{\left(1+\xi_{\tau'}\right)}{2}  \zeta_{Y_{\tau}}  \zeta_{Y_{\tau'}}  } \right \rangle \,.
\end{align}
Note that we are allowed to perform averaging over $\xi_{\tau'}$ directly, due to the fact that both $Y_{\tau'}$ and $Y_{\tau}$ are  statistically independent 
of a random variable $\xi_{\tau'}$. Indeed,  $Y_{\tau'}$ depends on $\xi_{\tau'-1}$, $\xi_{\tau'-2}$, $\ldots, \xi_0$, while $Y_{\tau}$, with $\tau < \tau'$, depends on $\xi_{\tau-1}$, $\xi_{\tau'-2}$, $\ldots, \xi_0$. 

Note that only the product $\zeta_{Y_{\tau}}  \zeta_{Y_{\tau'}} $ is dependent on random convection flows.  
Averaging this product over quenched disorder, we find that, in virtue of the definition in eq. \eqref{corr}, 
expression \eqref{mu} takes the form
\begin{align}
\label{zu}
  \overline{\delta_{\xi_{\tau},1}  \delta_{Y_{\tau},Y_{\tau'}}} \,,
\end{align}
i.e., it is an averaged over thermal noises product of the indicator functions of two events: a) $Y_{\tau+1} = Y_{\tau}$ and b) $Y_{\tau} = Y_{\tau+1} = Y_{\tau'}$. As a consequence, the expression \ref{zu} is the joint probability $P(Y_{\tau}|t=\tau'; Y_{\tau}|t=\tau+1;Y_{\tau}|t=\tau)$ of the events that 
the TP trajectory $Y_t$, with $Y_{t=0}=0$,  
a) paused at its (unspecified) position at $t = \tau$ and b) 
returned at time moment $t=\tau'$ to the position it occupied at $t =  \tau$.

The probability $P(Y_{\tau}|t=\tau'; Y_{\tau}|t=\tau+1;Y_{\tau}|t=\tau)$  decouples into the product of the probability that the trajectory $Y_t$ appeared at an unspecified position $Y_{\tau}$ at time moment $t = \tau$, which equals unity since averaging over 
$\xi_{k}$ with $k \in (0,\tau-1)$ implies averaging over all possible $Y_{\tau}$; the probability that $Y_t$
paused at $t = \tau$, which equals $1/2$; and  the probability 
that $Y_t$ returned to $Y_{\tau} = Y_{\tau+1}$ within $\tau'- \tau -1$ steps. 
Making a plausible 
assumption that
the dynamics, at least in the asymptotic limit $t \to \infty$, does not depend of the starting point,  we 
thus find
that the expression \eqref{zu} reduces to
\begin{align}
\label{zuz}
\frac{1}{2} P_{\tau' - \tau -1}\left(Y=0\right) \,,
\end{align}
where 
$P_{\tau' - \tau -1}\left(Y=0\right)$ is the probability that the $y$-component of the TP trajectory  
returns to $Y=0$, (not necessarily for the first time), on the $(\tau'-\tau-1)$-th step. Here, $P_{t}\left(Y\right)$ ($P_{t}\left(X\right)$) is a marginal distribution obtained from the full probability distribution function $P_t(X,Y)$ of finding the TP at site $(X,Y)$ at time moment $t$ by summing the latter over all $X$ ($Y$), that is 
\begin{align}
\label{marginal}
P_{t}\left(Y\right) = \sum_{X=-\infty}^{\infty} P_t(X,Y) \,, \,\,\, P_{t}\left(X\right) = \sum_{Y=-\infty}^{\infty} P_t(X,Y) \,.
\end{align}

\begin{figure}[htb!]
\begin{center}
\includegraphics[width=1.00\hsize]{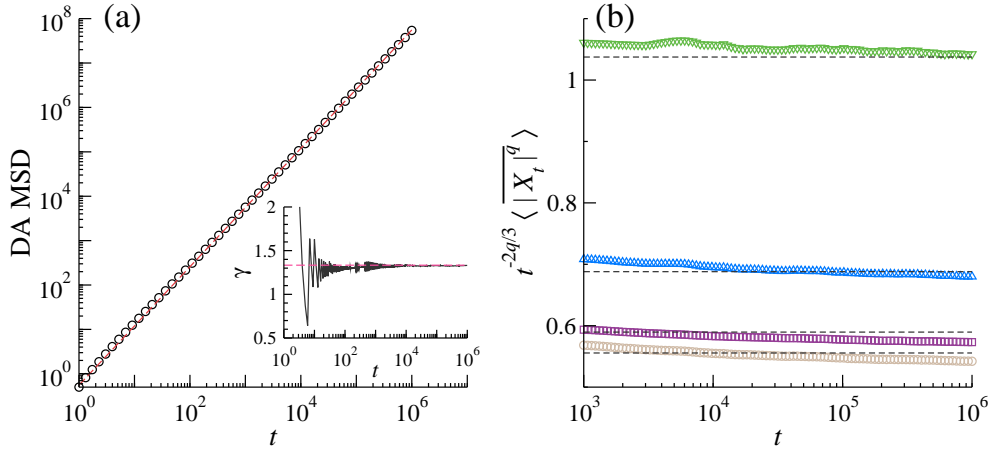}
\end{center}
\caption{{\em Disorder-averaged mean-squared displacement of the TP and higher moments of $X(t)$}.  Panel (a). Disorder-averaged mean-squared displacement $<\overline{X_t^2}>$ of the TP in absence of the dynamical environment (LG particles). The open circles depict our numerical results, while the dashed (red) line indicates 
the prediction $< \overline{X^2_t}> = m_2 \, t^{4/3}$ with $m_2 = 0.556$ (see eqs. \eqref{moments} and \eqref{mq}). 
A super-diffusive behaviour sets in from rather short times and the diffusive transient (see the first term in the rhs of eq. \eqref{kk}) is not observed.
The inset displays the rate of a convergence of the time-dependent dynamical exponent $\gamma_t$, eq. \eqref{gammat}, to its asymptotic value $4/3$. Panel (b). Reduced moments $<\overline{|X_t|^q}>\big/t^{2 q/3}$ as functions of time. The dashed lines (from top to bottom) correspond to $m_4 = 1.038$, $m_3 = 0.687$, $m_1 = 0.590$ and $m_2 = 0.556$ (see eq. \eqref{mq}).
}
\label{fig2}
\end{figure}

\begin{figure}[htb!]
\begin{center}
\includegraphics[width=1.0\hsize]{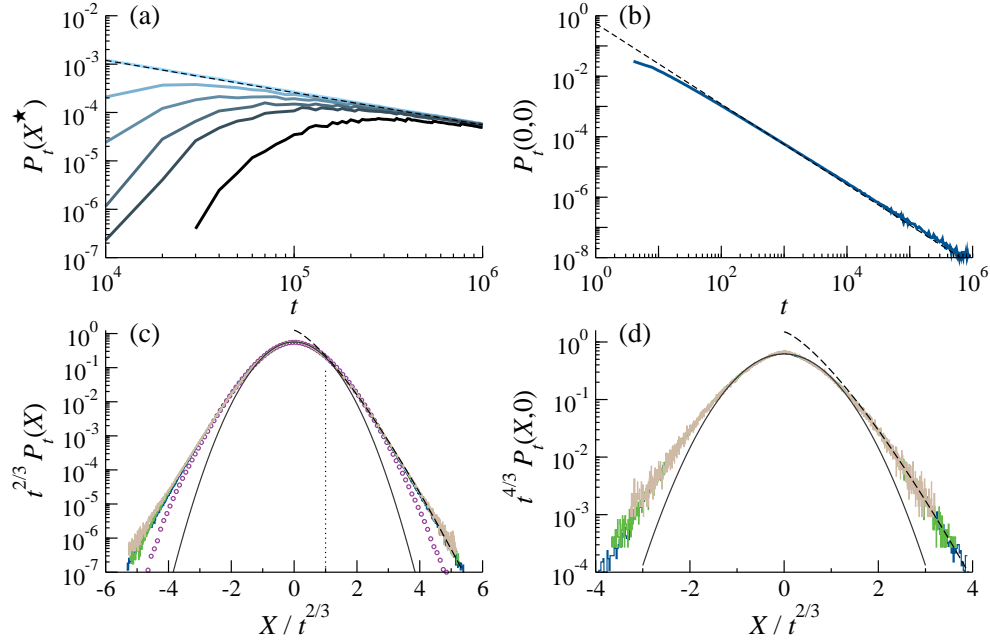}
\end{center}
\caption{{\em Probability distribution of the TP position}. Panel (a). Temporal evolution of the marginal distribution $P_t(X=X^{\star})$, eq. \eqref{marginal}, 
for 
six fixed values
of $X^{\star}=0, 60, 1000, 1400, 1800$ and $3000$  (solid curves from top to bottom  with lighter curves corresponding to smaller values of $X^{\star}$). The 
dashed line denotes the power-law $A/t^{2/3}$ with $A \approx  0.568$. Note that $P_t(0) \simeq A/t^{2/3}$
 provides a very accurate estimate for $P_t(0)$ starting from rather short times; $P_t(X=X^{\star})$ for 
 $X^{\star} > 0$ converges ultimately to $P_t(0)$.
 Panel (b). The probability $P_t(0,0)$ of being at the origin at time moment $t$. Thick blue curve presents the numerical data. A dashed line is a power-law $A'/t^{4/3}$ with $A' \approx 0.555$. It provides a fairly good estimate for the numerical data starting from $t \geq 10^2$. Panel (c). The marginal distribution $P_t(X)$, multiplied by $t^{2/3}$, is plotted as a function of the scaled variable $u = X/t^{2/3}$. Panel (d). The full distribution $P_t(X,Y)$ with $Y=0$, multiplied by $t^{4/3}$, is plotted as a function of the scaled variable $u = X/t^{2/3}$.
In panels (c) and (d) the histograms show the results of numerical simulations: light brown, green and blue colours correspond to the numerical data for $t= 10^4$, $4 \times 10^4$ and  $t = 10^5$, respectively; thin solid curves are the Gaussian function $A \exp(- a u^2)$, and  the dashed curves - a stretched-exponential function of the form $B \exp(- b |u|^{4/3})$.
Vertical dotted line in panel (c) is a guide to an eye which indicates the crossover value $u=1$ between the two asymptotic regimes.
Purple circles in the panel (c) depict our data for a shorter time -  $t=10^3$. A deviation 
from the stretched-exponential form signifies that the anomalous tails of $P_t(X)$ appear only for  sufficiently large values of $t$.
}
\label{fig3}
\end{figure}

Summing up the presented above reasonings, we arrive at the following representation of the DA MSD:
\begin{align}
\label{kk}
\left \langle \overline{X_{t}^2} \right \rangle \sim \frac{t}{2} +  \frac{1}{2} \sum_{\tau=0}^{t-2} \sum_{\tau' = \tau+1}^{t-1} P_{\tau' - \tau -1}\left(Y=0\right) \,.
\end{align}
Further on, the probability $P_{\tau' - \tau-1}\left(Y=0\right)$ is evidently 
a decreasing function of the difference $\tau'-\tau-1$. 
Very general arguments (see also the numerical results presented in Fig. \ref{fig3}, panel (a)), suggest that  
$P_{\tau' - \tau-1}\left(Y=0\right)$ decays as a power-law :
 \begin{align}
 \label{powerlaw}
P_{\tau' - \tau-1}\left(Y=0\right) \sim \frac{A}{(\tau'-\tau-1)^{\gamma/2}}
\end{align} 
 in the limit $(\tau'-\tau-1) \to \infty$, where $A$ is the amplitude and $\gamma$ is the dynamical exponent, both to be defined. Supposing that 
 $\gamma < 2$ ($\gamma = 2$ corresponds to ballistic motion), we expect that 
both the 
 inner sum (over $\tau'$) and the outer one (over $\tau$) in eq. \eqref{kk} will be dominated by the upper summation limit. 
As a consequence, in the large-$t$ limit
\begin{align}
\label{dsum}
\frac{1}{2}   \sum_{\tau=0}^{t-2} \sum_{\tau' = \tau+1}^{t-1} P_{\tau'- \tau - 1}\left(Y=0\right) \sim \frac{A t^{2 - \gamma/2}}{2 (1 - \gamma/2)(2 - \gamma/2)} \,,
\end{align}
and hence, in the large-$t$ limit the expression \eqref{kk} attains the form
\begin{align}
\left \langle \overline{X_t^2} \right \rangle \sim \frac{t}{2} + \frac{A t^{2 - \gamma/2}}{2  (1 - \gamma/2)(2 - \gamma/2)} \,.
\end{align}
In line with the arguments presented in Refs. \cite{redner,ledoussal2}, we  
recall that the dynamical exponent $\gamma$ defines the characteristic extent of the trajectory $Y_{t}$; that being, $<\overline{Y^2_{t}}> = m_2 \, t^{\gamma}$, where $m_2$ is as yet unknown proportionality factor.
By symmetry, one expects thus that the DA MSD along the $x$-axis, i.e., $<\overline{X^2_{t}}>$, obeys exactly the same law, which entails the following closure relation :
\begin{align}
\label{rel}
 m_2 \, t^{\gamma} \sim \frac{t}{2} + \frac{A t^{2 - \gamma/2}}{2 (1 - \gamma/2)(2 - \gamma/2)} \,.
\end{align}
Inspecting the behaviour of the latter expression in the limit $t\to \infty$, we infer that the contribution of the first term in the rhs of eq. \eqref{rel} becomes negligible in the limit $t \to \infty$, so that the dominant contribution is provided by the second term.
Comparing the power-law on the left-hand-side (lhs) of eq. \eqref{rel} with the second term on the rhs of this equation, we find that
the exponent $\gamma$ obeys
\begin{align}
\gamma = 2 - \gamma/2 \,,
\end{align}
which yields $\gamma = 4/3$ - the value which has been previously conjectured and verified numerically in Refs. \cite{redner,ledoussal2}.  

Therefore, our reasonings correctly reproduce the value of the dynamical exponent $\gamma$. However, 
inferring a numerical value of the prefactor $m_2$ from eq. \eqref{rel}, (which predicts $m_2 \sim 9 A/8$), should lead to a somewhat higher $m_2$ than the actual one, because the rhs in eq. \eqref{dsum} evidently overestimates the value of the double sum in the lhs of this equation. 
The point is that the algebraic
form in eq. \eqref{powerlaw} is only valid 
for such realisations of the TP trajectories, for which
 the sum of the number of jumps and of the number of the pausing events is even.
Otherwise, $P_{\tau'-\tau-1}(Y=0)$  is exactly equal to zero.
 As a consequence, eq. \eqref{rel}  overestimates $m_2$.
 
Lastly, we note that a  similar type of arguments was invoked to characterise a
decay of the number of tree-like clusters with a growing pattern height
in a process of ballistic deposition of sticky particles on a line   \cite{sergei}. 
Both the decay 
and the ensuing thinning of the forest of such clusters appear to be controlled by 
a  random wandering of the inter-cluster boundaries with the super-diffusive exponent $\gamma = 4/3$.

In Fig. \ref{fig2}, panel (a), we 
present numerical results (open circles) describing 
the time evolution of the DA MSD of a single TP. 
The dashed line indicates the 
super-diffusive power-law behaviour of the form $<\overline{X^2_t}> \sim m_2 \,  t^{4/3}$, with $m_2 = 0.556$. This estimate of 
$m_2$ 
is based on the fitting of the full probability distribution, which is discussed below in subsection \ref{pdf}. We observe that 
the super-diffusive behaviour sets in from rather early times and the transient diffusive law, as predicted by the first term in the rhs of eq. \eqref{rel}, is not observed.
Next, the inset in the panel (a) illustrates the convergence of the dynamical exponent
$\gamma_t$, defined by
\begin{align}
\label{gammat}
\gamma_t = \dfrac{\left(\ln\left(\langle \overline{X_t^2}\rangle\right) - \ln\left(\langle \overline{X_{t^{z}}^2}\rangle\right)\right)}{(1-z) \ln(t)}\,,
\end{align}
to its asymptotic value $4/3$. Such a representation of $\gamma_t$ (as compared to the standardly used one, $\gamma_t = \ln\left(\langle \overline{X_t^2}\rangle\right)/\ln(t)$) is particularly well-suited for a numerical analysis of 
 the dynamical exponent in an expected power-law dependence on time with an unknown numerical prefactor, since the latter cancels out automatically. 
In eq. \eqref{gammat} the parameter 
$z$ is a trial exponent, $0 < z <1$, which rescales time in the second term; in principle, $z$ can be chosen rather arbitrarily;  we use $z = 0.9$. We also observe that $\gamma_t$ converges to its asymptotic value very rapidly, in line with the behaviour of the DA MSD.
In panel (b) of Fig. \ref{fig2} we 
plot the reduced moments
 $<\overline{|X|^q_t}>/t^{2 q/3}$ for $q = 1, 2, 3$ and $4$ as functions of time. We observe that the reduced moments saturate as some constant values $m_q$ as time progresses, indicating that the moments themselves obey $<\overline{|X|^q_t}> = m_q  \, t^{2 q/3}$ (see eq. \eqref{moments}). Here, the dashed lines
  indicate our estimates for the values of the numerical prefactors $m_q$ (see eq. \eqref{mq}). 
 
\subsection{Probability distribution and moments of arbitrary order}
\label{pdf}

In Fig. \ref{fig3} we depict different facets of the numerically evaluated 
full probability distribution $P_t(X,Y)$ and of the marginal distribution  
$P_t(X)$, (see eq. \eqref{marginal}). 
Panel (a) presents the time evolution of $P_t(X=X^{\star})$ for six fixed values of 
$X^{\star}$: $X^{\star}=0, 60, 1000, 1400, 1800$ and $3000$ (curves from top to bottom, with lighter colours corresponding to smaller values of $X^{\star}$).
 Our numerical results show that, unequivocally, $P_t(0)$ obeys a power-law of the form $P_t(0) \simeq A/t^{2/3}$, which is fully in line with our above analysis.  The decay amplitude is defined with a good accuracy
by $A \approx  0.568$.   Moreover, comparing our numerical results with the form $P_t(0) \simeq A/t^{2/3}$,
we conclude that the latter 
 provides a very accurate estimate for $P_t(0)$ starting from rather short times - the dashed line representing $A/t^{2/3}$ and the numerical data (light blue curve) are almost indistinguishable.
In turn, $P_t(X =X^{\star})$ for $X^{\star}=60, 1000, 1400, 1800$ and $3000$
converges ultimately to $P_t(0) \simeq A/t^{2/3}$, which is, of course, not an unexpected behaviour.
 The panel (b) presents the time evolution of $P_t(0,0)$ - the probability of being at the origin at time moment $t$. We observe that the power-law form $P_t(0,0) \simeq A'/t^{4/3}$ (with $A' \approx 0.555$) describes the numerical data fairly well. Note also that this form implies  
 that a random walk on a random Manhattan lattice is not certain to return to the origin.

Further on, in Fig. \ref{fig3}, panels (c) and (d), we plot $t^{2/3} P_t(X)$ and  $t^{4/3} P_t(X,Y)$ with $Y=0$ as functions of the scaled variable $u = X/t^{2/3}$.
The data collapse evidenced by our numerical results for both the central part of the distribution and for its tails, suggests, again  
rather unequivocally, that the marginal distribution $P_t(X)$ 
of the TP position along the $x$-axis at (sufficiently large) time $t$ has the following form:
\begin{equation}
\label{dist}
P_t(X) =  \frac{1}{t^{2/3}} \begin{cases} A \exp\left(- a u^2\right) &\mbox{for } |u| < 1 \\
B \exp\left(- b |u|^{4/3}\right) & \mbox{for } |u| >  1 \end{cases}
\end{equation}
where $B \approx 1.249$, $a \approx 1.049$ and $b \approx 1.730$. We observe, as well, that
the full distribution $P_t(X,Y)$ (with $Y=0$) exhibits essentially the same functional behaviour as a function of $u$, (see Fig. \ref{fig3}, panel (d)), as the marginal distribution $P_t(X)$ and only the values of the parameters are slightly different. 
We therefore conclude that a) the central part of both distributions is a Gaussian, with the variance which grows super-diffusively with $t$, and b) the tails of both distributions deviate from a Gaussian and have a form $\sim \exp( - |u|^{4/3})$, i.e., are "heavier" than a Gaussian. The presence of such tails also manifests itself in the anomalously high 
asymptotic value $\approx 3.5$ attained by the kurtosis of the marginal distribution $P_t(X)$ (see the dashed curve in Fig. \ref{fig7}, panel (d)). Recall that the kurtosis of a Gaussian distribution is equal to $3$.

We note that the large-$u$ tail of $P_t(X)$ and $P_t(X,Y=0)$ has a very different form, as compared to the prediction made in Refs. \cite{redner,ledoussal2}.  Assuming the validity of the usual relation between the shape exponent $\delta$ and the dynamical exponent $\gamma$, $\delta = 1/(1-\gamma)$, it was conjectured that the shape exponent
should be $\delta = 3$. Our data shows that this is not the case and, 
surprisingly enough, the distribution
 in the second line in eq. \eqref{dist} has exactly the same shape exponent $\delta = 4/3$ as the one appearing in the MdM model with random layered convection flows (see Refs. \cite{redner,ledoussal2,ledoussal3}).

Capitalising on the expression in eq. \eqref{dist}, 
 we estimate the behaviour of the moments of $P_t(X)$ of arbitrary order $q$. 
 Multiplying both sides of eq. \eqref{dist} by $|X|^q$, changing the integration variable for $u = X/t^{2/3}$, and integrating the expression in the first line over $u \in (-1,1)$ and  in the second line - over $u \in (-\infty,-1)$ and $(1,\infty)$, we get
\begin{align}
\label{moments}
\left \langle \overline{|X_t|^q} \right \rangle = m_q \, t^{2 q/3}, 
\end{align}
with
\begin{align}
\label{mq}
m_q \approx \frac{A}{a^{(q+1)/2}} \Big(\Gamma(q+1) - \Gamma(q+1,a)\Big) + \frac{3 B}{2 b^{3 (q+1)/4}} \Gamma\left(\frac{3(q + 1)}{4},b\right) \,,
\end{align}
where $\Gamma(a,b)$ is the incomplete Gamma-function. Note that here we discard the transient region between two asymptotic regimes, supposing that the second regime is valid starting from $|u| = 1$. This is, of course, not true and hence, $m_q$ in eq. \eqref{mq} \textit{overestimates} the actual value of the numerical prefactor $m_q$ in eq. \eqref{moments}. We however believe that such an estimate is quite plausible and would not incur any significant error.
The plot of the numerical results for the first four moments together with the estimates for $m_q$ presented
in Fig. \ref{fig2}, panel (b),
shows that it is indeed the case. 

We close this subsection with two following remarks: a)  the value of $m_2$ deduced from eq. \eqref{rel}, i.e., $m_2 \approx 9 A/8 \approx 0.639$, slightly overestimates the value of $m_2$ obtained from eq. \eqref{mq}, i.e., 
 $m_2 = 0.556$. This is completely in line with our argument that the second term in the rhs in eq. \eqref{rel} provides an upper bound on the actual value of $m_2$. b) 
For the kurtosis $\kappa$ of the marginal distribution $P_t(X)$, i.e.,
\begin{align}
\label{kappa}
\kappa = \left \langle \overline{X_t^4} \right  \rangle \Big/ \left \langle \overline{X_t^2} \right \rangle^2 \,,
\end{align} 
we find from eqs. \eqref{moments} and \eqref{mq} that $\kappa = m_4/(m_2)^2 \approx 3.360$, 
which value favourably agrees with $\kappa \approx 3.5$ deduced from our numerical simulations
(see Fig. \ref{fig7}, panel (d)).

\begin{figure}[htb!]
\begin{center}
\includegraphics[width=1.00\hsize]{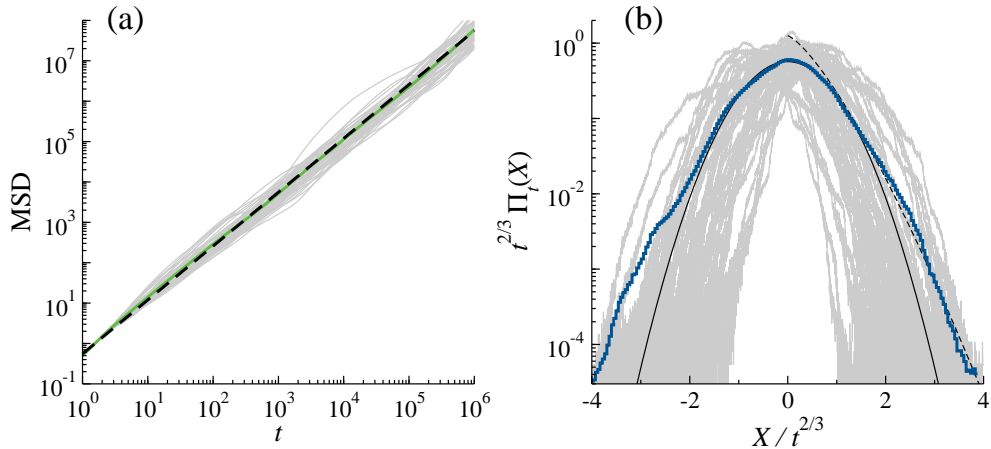}
\end{center}
\caption{{\em Sample-to-sample fluctuations of the TP trajectories}. Panel (a). MSD 
$\overline{X_t^2}$ for fifty (chosen at random) realisations of random convection flows plotted as a function of time.  The dashed line represents the DA MSD $\langle \overline{X_t^2} \rangle = 0.556 \, t^{4/3}$ (see Fig. \ref{fig2}), while the green line indicates the MSD averaged over $50$ realisations of disorder.
Panel (b). Realisation-dependent probability distribution function ${\Pi}_t(X)$ of the TP position along the $x$-axis, multiplied by $t^{2/3}$, is plotted versus a scaled variable $X/t^{2/3}$. We present $50$ (thin grey) curves  corresponding to $50$ fixed realisations of frozen convection flows.
Thin solid and dashed curves depict the Gaussian function $A \exp(- a u^2)$ and a stretched-exponential function of the form $B \exp(- b |u|^{4/3})$, 
 respectively, with the parameters $A, a, B$ and $ b$ as defined in Fig. \ref{fig3}. Blue zigzag curve represents ${\Pi}_t(X)$ averaged over $50$ realisations of random convection flows. }
\label{fig4}
\end{figure}

\subsection{Sample-to-sample fluctuations}

Up to the present moment we discussed only the averaged behaviour - both over thermal histories and over realisations of  random convection flows. However, a legitimate question is how the pertinent parameters themselves vary from a realisation to a realisation of the latter pattern. 
This question has been first addressed in Refs. \cite{redner,ledoussal2} for the MdM model with random layered flows and significant sample-to-sample fluctuations have been predicted. However, for the dynamics on a random Manhattan lattice this issue has not been analysed and we concentrate on it below,  focusing on the mean-squared (averaged over thermal histories only) displacement $\overline{X_t^2}$ of the TP on a given pattern of arrows  as well as on the corresponding probability distribution function ${\Pi}_t(X)$ of its position $X$ along the $x$-axis. Recall that 
$P_t(X) = \langle {\Pi}_t(X)\rangle$. 

In Fig. \ref{fig4}, panel (a),  we depict 
the corresponding MSD $\overline{X^2_t}$ (averaging over thermal histories is performed over $2 \times 10^7$ realisations of trajectories, for a given realisation of convection flows) for $50$ realisations  of disorder
as functions of time. We do indeed 
observe some scatter in the values of the prefactor $m_2$, which here is a random variable dependent on a particular realisation of disorder. On the other hand, the amplitude of fluctuations does not seem to be very significant and all the curves concentrate essentially around  the DA MSD $\langle \overline{X_t^2} \rangle = 0.556 \, t^{4/3}$ (see Fig. \ref{fig2}). Moreover, we  realise that the MSD averaged over $50$ realisations  
of disorder only, appears to be fairly close to the DA MSD evaluated using an ample statistical sample; recall that in subsection \ref{pdf} we used $2 \times 10^5$ realisations of disorder in order to perform averaging over random convection flows. On contrary, the sample-to-sample fluctuations do affect in a  significant way the shape of the distribution function ${\Pi}_t(X)$, both in the central part and especially in the region of anomalous tails, for which the numerical data looks quite nebulous.  However, it is quite surprising to realise that being averaged over just $50$ realisations of disorder, ${\Pi}_t(X)$ gets rather close to $P_t(X)$ depicted in Fig. \ref{fig3} (see thin solid and dashed curves in Fig. \ref{fig4}), which again was evaluated using a much bigger statistical sample. Here, an agreement between 
the blue zigzag curve and the thin black solid line looks nearly perfect
within the central, Gaussian part of the distribution, while also for the tails it shows a rather convincing agreement. Therefore, we may conclude that sample-to-sample fluctuations are essentially less important  for a random motion on a random Manhattan lattice than in the MdM model with layered random flows \cite{redner,ledoussal2}.

\subsection{Spectral analysis of the TP trajectories}

Complementary information about anomalous 
diffusion of the TP can be inferred from
 the so-called single-trajectory power spectral density $S(T,f)$, where $T$ is the observation time and $f$ is the 
 frequency (see, e.g., Ref. \cite{we1} for more details). This property is a random, realisation-dependent variable, 
 parametrised by $f$ and $T$.
 Note that in the model at hand, it depends on both a given pattern of arrows in a random Manhattan lattice and on a
 given realisation of a thermal history.  
 For an integer-valued $X_t$,  $S(T,f)$ is defined as 
 \begin{align}
 \label{power}
 S(T,f) = \frac{1}{T} \left|\sum_{t=0}^T e^{i f t} \, X_t  \right|^2 \,,
 \end{align}
 and hence, is a periodic function of $f T$ with the prime period $2 \pi T$. 
 
In a standard text-book analysis, one considers
the ensemble-averaged (and also disorder-averaged for our case) value of the random variable in eq. \eqref{power}, i.e. its first moment :
 \begin{align}
 \label{mul}
 \mu(T,f) = \left \langle \overline{S(T,f)}\right\rangle =  \frac{1}{T} \sum^T_{t_1, t_2=0} \cos\left(f \left(t_1 - t_2\right)\right) \left \langle \overline{ X_{t_1} X_{t_2}}\right\rangle\,, 
 \end{align} 
 which probes the frequency-dependence of the Fourier-transformed covariance function of $X_t$.
 Moreover,  one also takes formally 
 the limit of an infinitely long observation time, i.e., sets $T = \infty$. 
 We will demonstrate below that, although one has indeed to consider very large values of $T$ in order to extract a meaningful information about the $f$-dependence of the power spectral density in eq. \eqref{mul}), taking the formal limit $T = \infty$
  in our case renders such a standard definition meaningless.
 
 \begin{figure}[htb!]
\begin{center}
\includegraphics[width=1.00\hsize]{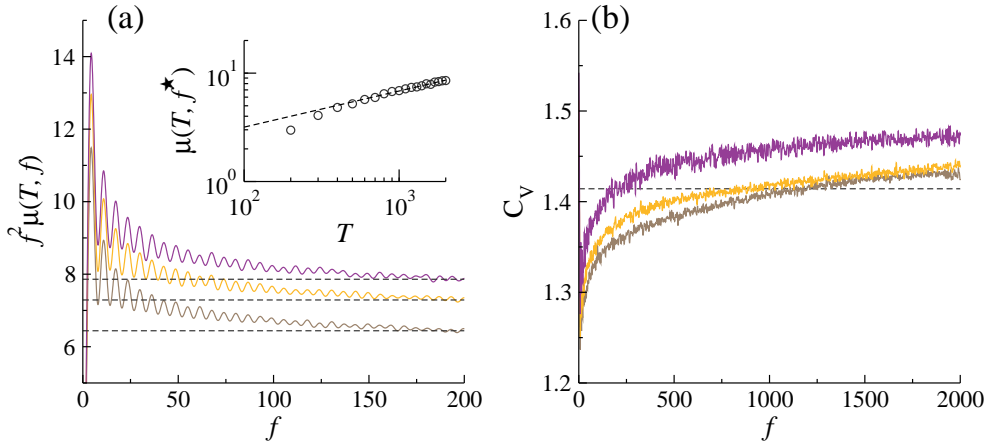}
\end{center}
\caption{{\em Spectral properties of the TP trajectories}. Panel (a). Ensemble- and disorder-averaged power spectral density $\mu(T,f)$, eq.
\eqref{mul},  multiplied by $f^2$,  as a function of the 
 frequency $f$ for three values of the observation time $T$ (from top to bottom $T = 4800, 2400$ and $1200$).  The inset displays the \textit{ageing} behaviour of $\mu(T,f=f^{\star})$ as a function of the observation time $T$ for a fixed frequency $f^{\star} = 5 \times 10^2$, and evidences the dependence $\mu(T,f=f^{\star}) \sim T^{1/3}$. Panel (b). The coefficient $C_v$ of variation, eq. \eqref{cv}, of the distribution of a random variable $S(T,f)$, eq. \eqref{power}, as a function of the frequency $f$ for three values of the observation time $T$ (the same as in panel (a)). The dashed line indicates $C_v = \sqrt{2}$ - a value specific to a super-diffusive fractional Brownian motion (with arbitrary Hurst index in the interval $1/2 < H < 1$ 
 and hence, with $\gamma$ in the interval $1 < \gamma < 2$).}
\label{fig5}
\end{figure}

In Fig. \ref{fig5} we present the results of a numerical analysis of the functional form of $ \mu(T,f)$, which reveals two rather surprising features. First, it appears that  $\mu(T,f)$ is \textit{ageing}, i.e., its amplitude is dependent on the observation time $T$,   $\mu(T,f) \sim T^{1/3}$, which is demonstrated in the inset to this figure.  As a consequence, setting $T = \infty$ is meaningless. 
Second, we observe that $f^2 \mu(T,f) $, (for three fixed values of $T$), approaches constant $f$-independent values for sufficiently large $f$. This signifies that $\mu(T,f) \sim 1/f^2$, i.e., it exhibits exactly the same $f$-dependence 
as the power spectral density of a standard Brownian motion (see, e.g., Ref. \cite{we1} and references therein), although the process under study is clearly not a Brownian motion. Therefore, fixing $T$ and focussing only on the $f$-dependence
of  $\mu(T,f)$ garnered from numerical simulations, one can be led to an erroneous conclusion that the observed process is a Brownian motion. As an actual fact, this is precisely the $T$-dependence of  $\mu(T,f)$ which helps to realise that this is not the case (see also Ref. \cite{Krapf_2019}).

We note that such a "deceptive" $f$-dependence has been previously reported for the running maximum of Brownian motion \cite{carlos}, diffusion in a periodic Sinai disorder \cite{enzo}, diffusion with stochastic reset \cite{satya} and also for a variety of diffusing diffusivity models \cite{we}. 
Further on, the law $\mu(T,f) \sim T^{1/3}/f^2$ was observed for other super-diffusive processes, such as a fractional Brownian motion with the Hurst index $H = 2/3$ (i.e., $\gamma=4/3$) \cite{Krapf_2019} or a super-diffusive scaled Brownian motion $Z_t$ described by the Langevin equation $\dot{Z_t} = t^{1/6} \zeta_t$ \cite{sposini}, with $\zeta_t$ being a Gaussian white-noise with zero mean. This latter process also produces a super-diffusive motion with $\gamma = 4/3$, 
suggesting that the law  $\mu(T,f) \sim T^{1/3}/f^2$ might be a generic feature of processes with $\gamma = 4/3$. We note parenthetically that this questions
the robustness of the textbook approach, based solely on the evaluation of $\mu(T,f)$, which 
 cannot distinguish between these three distinctly different random processes.

The difference between  these processes becomes apparent, however, when one considers higher-order moments of $S(T,f)$, e.g., its variance. In particular, one may focus on 
the 
coefficient of variation $C_v$, which is defined by
\begin{align}
\label{cv}
C_v = \sigma(T,f)\Big/\mu(T,f)  = \sqrt{ \left(\left \langle \overline{S^2(T,f)}\right\rangle -  \left \langle \overline{S(T,f)}\right\rangle^2\right)} \Big/\mu(T,f) \,,
\end{align}
where $\sigma(T,f)$ is the standard deviation of a random variable $S(T,f)$.
This characteristic parameter 
shows a completely different behaviour as a function of $f$ for a super-diffusive fractional Brownian motion and a super-diffusive scaled Brownian motion. For the former $C_v$ approaches for sufficiently large $T $ and $f$ 
a universal (i.e., regardless of the actual value of  $H > 1/2$), time $T$-independent  constant 
value $\sqrt{2}$ \cite{Krapf_2019}, while for the latter - a universal time $T$-independent 
constant value $\sqrt{5}/2$  \cite{sposini}, 
the same which is observed for a standard Brownian motion \cite{we1}.

To this end, we have studied via numerical simulations the frequency and the observation time dependence of $C_v$ for the TP random motion on a random Manhattan lattice. This dependence is presented 
in Fig. \ref{fig5}, panel (b), in which we plot the coefficient
 $C_v$ of variation of a single-trajectory power spectral density as a function of $f$ for three values of $T$.  We observe that 
$C_v$ tends to a higher than $\sqrt{2}$ value as frequency increases. Moreover, $C_v$ is clearly ageing, i.e., its limiting behaviour is dependent on the observation time.  In conclusion, we observe a behaviour of $C_v$ 
which is markedly different from the two above mentioned examples of super-diffusion with $\gamma = 4/3$.

\section{Tracer particle dynamics on a populated random Manhattan lattice}
\label{particles}

\begin{figure}[htb!]
\begin{center}
\includegraphics[width=1.00\hsize]{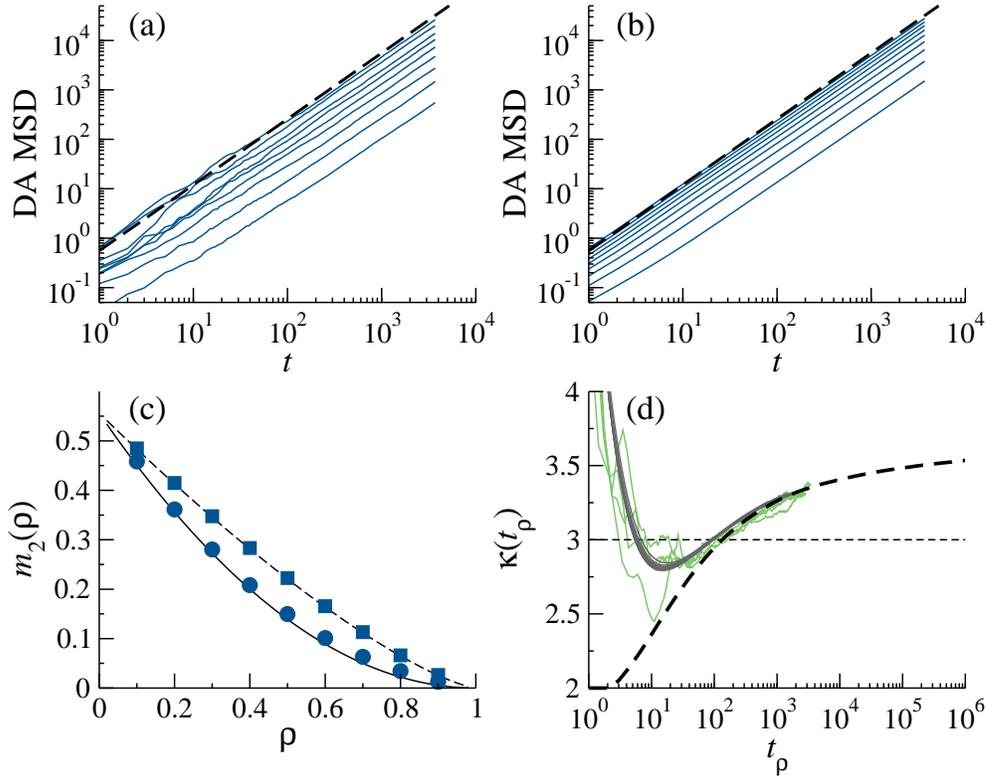}
\end{center}
\caption{{\em DA MSD of the TP,  prefactor $m_2(\rho)$ and the kurtosis for Model A and Model B}. Panel (a). 
The DA MSD for the Model A as a function of time for different  densities of the LG particles. Curves from top to bottom correspond to $\rho = 0.1, 0.2, \ldots, 0.9$. Black dashed line depicts the DA MSD $\langle \overline{X_t^2} \rangle = 0.556 \, t^{4/3}$ of a single TP (in absence of the LG particles, i.e., for $\rho = 0$). Panel (b). The DA MSD as a function of time for Model B. Curves from top to bottom correspond to $\rho = 0.1, 0.2, \ldots, 0.9$ and the black dashed line corresponds to $\rho =0$. Panel (c). Numerical prefactor $m_2(\rho)$ in the law $\langle \overline{X_t^2} \rangle =  m_2(\rho) \, t^{4/3}$
for Model A and Model B. Symbols present the results of numerical simulations: filled circles (Model A) and filled squares (Model B). Thin solid and dashed  lines are fits to the numerical data: solid  line is $m_2(\rho) = 0.556 \, (1 - \rho)^{2}$ (Model A)
and the dashed line is $m_2(\rho) = 0.556 \, (1 - \rho)^{4/3}$ (Model B). Panel (d). The kurtosis $\kappa$ of the distribution $P_t(X)$, (defined in eq. \eqref{kappa}), as a function of a rescaled time $t_{\rho}$. Dashed line depicts the kurtosis of $P_t(X)$ of a single TP, here, $t_{\rho} = t$; noisy green curves present the time evolution of the kurtosis for Model A for $\rho = 0.1, 0.2, 0.3$ and $0.4$, here $t_{\rho} = (1-\rho)^{3/2} t$; and eventually the
grey curves, which collapse on a single master curve (dashed line), depict the kurtosis of the corresponding distribution $P_t(X)$ for Model B with $\rho=0.1, 0.2, \ldots, 0.9$, here, $t_{\rho} = (1 - \rho) t$.}
\label{fig6}
\end{figure}

In this last section we discuss the results of 
 the numerical analysis of the TP dynamics on a crowded Manhattan lattice. In Fig. \ref{fig6} we  present the DA MSD of the TP for Model A and Model B,  prefactor $m_2(\rho)$ in the super-diffusive law $\langle \overline{X_t^2} \rangle = m_2(\rho) \, t^{4/3}$ as a function of the density $\rho$ of the LG particles, 
and also the kurtosis of the distribution $P_t(X)$ 
for Model A and Model B.  
First of all, we realise that for both models the DA MSD obeys the same super-diffusive law $\langle \overline{X_t^2} \rangle \sim t^{4/3}$, for any density of the LG particles. Prefactor $m_2(\rho)$ depends, of course, on the density of the LG particles and their dynamics; indeed, $m_2(\rho)$ shows apparently different dependences on $\rho$ for Model A and Model B. For Model A, in which the TP follows random convection flows while the LG particle perform constrained random walks, 
 this dependence is most strong and is rather close to a parabolic law $m_2(\rho) = 0.556 \, (1 - \rho)^{2}$ (thin solid curve in panel (c)). This parabolic law very accurately describes the actual dependence of $m_2(\rho)$ on $\rho$ for $\rho < 1/2$. For higher densities, however, some deviations are clearly seen, although such a discrepancy can be also attributed to the lack of a large enough statistical sample. For Model B, in which all particles are identical and all perform a super-diffusive motion, the $\rho$-dependence of the prefactor  $m_2(\rho)$ is given by $m_2(\rho) = 0.556 \, (1 - \rho)^{4/3}$ (thin dashed curve in panel (c)). This law agrees fairly well with the numerical data for any value of the LG particles density. In turn,  as we have mentioned above, such a dependence implies 
 that the system is perfectly stirred and the time variable $t$ gets merely rescaled by the fraction of successful jump events of the TP, i.e., by $(1 - \rho)$, as one can expect from simple mean-field-type arguments.  Arguably, this expression for $m_2(\rho)$ is exact.

Next, in Fig. \ref{fig6}, panel (d), we depict the time evolution of the kurtosis of the distribution $P_t(X)$ in case of a single TP, as well as for Model A and Model B. Thick dashed line corresponds to the kurtosis of $P_t(X)$ in case of a single TP. We observe that $
\kappa$ saturates as $t$ evolves at a constant value which is close to $3.5$, i.e., it exceeds the value $3$ specific to a Gaussian distribution and thus implies that $P_t(X)$ is not Gaussian. This happens, of course, due to the presence of anomalous slower-than-Gaussian tails, which we discussed in subsection \ref{pdf}. Next, noisy green curves depict our results for $\kappa$ in Model A, with the LG particles densities $\rho = 0.1, 0.2, 0.3$ and $\rho=0.4$, plotted versus a rescaled time variable $t_{\rho} = (1-\rho)^{3/2} t$. Although there is a significant scatter of these curves at short times, we notice that  for sufficiently large $t_{\rho}$ all these curve collapse on the dashed line representing a single TP case. For Model B 
we analyse the TP dynamics for a broader range of the LG particles densities; we considered  nine values of $\rho$, $\rho = 0.1, 0.2, \ldots, 0.9$. Here, the curves defining the evolution of $\kappa$ for different values of $\rho$,  merge altogether even for short times when plotted versus a rescaled time $t_{\rho} = (1-\rho) t$, and eventually approach the value of the kurtosis for a single TP case.
Such a behaviour signifies that for sufficiently large times $P_t(X)$ possesses some universal scaling properties for both Model A and Model B.

\begin{figure}[htb!]
\begin{center}
\includegraphics[width=1.00\hsize]{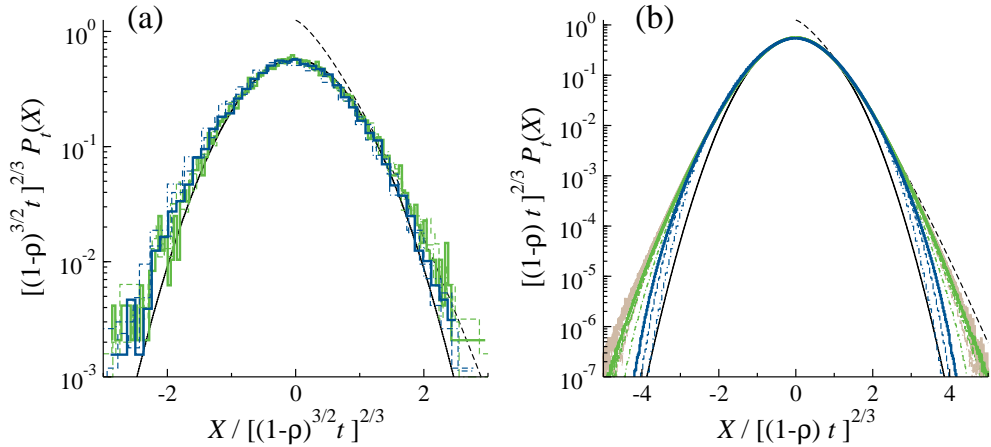}
\end{center}
\caption{{\em Probability distribution of the TP position along the $x$-axis on a populated random Manhattan lattice}. Panel (a). $P_t(X)$ multiplied by $(1-\rho) t^{2/3}$ is plotted as a function of a scaled variable $u_1= X/((1-\rho)^{3/2} t)^{2/3}$. Green zigzag curves correspond to $\rho = 0.2$, while the blue ones - to $\rho = 0.8$. Dot-dashed, dashed and thick solid zigzag curves depict the numerical data for $t = 10^2, 2 \times 10^3$ and $3 \times 10^3$, respectively.  
Thin solid curve is a Gaussian function $A \exp\left( - a u_1^2\right)$, while the thin dashed curve is an anomalous tail of the form $B \exp\left(-b |u_1|^{4/3}\right)$. Panel (b). $P_t(X)$ multiplied by $((1-\rho) t)^{2/3}$ is plotted as a function of a scaled variable $u_2= X/((1-\rho) t)^{2/3}$. Brown zigzag curves correspond to $\rho = 0.1$, green - to $\rho = 0.5$ and blue - to $\rho = 0.9$. 
Dot-dashed, dashed and thick solid zigzag curves correspond to the same values of $t$ as in panel (a). 
Thin solid curve depicts a Gaussian function $A \exp\left( - a u_2^2\right)$, while the thin dashed curve depicts an anomalous tail of the form $B \exp\left(-b |u_2|^{4/3}\right)$. 
Parameters $A, a, B$  and $b$ in both panels have the same values as in Fig. \ref{fig3}.}
\label{fig7}
\end{figure}

Lastly, in Fig. \ref{fig7} we present the numerical data for the distribution function $P_t(X)$ for Model A and Model B. 
We observe that for both models the central Gaussian part of the distribution is described with a very good accuracy by a Gaussian function:  
\begin{align}
P_t(X) = \frac{A}{t_{\rho}^{2/3}} \exp\left(- a \frac{X^2}{t_{\rho}^{4/3}}\right) \,,
\end{align}
where the choice of $t_{\rho}$ depends  on the model under study. For a single TP case, $t_{\rho} = t$.
The lack of a big statistical sample does not permit  to make completely conclusive statements about the tails of the distribution. We notice, however, that such tails are definitely present and a departure of the distribution from a purely Gaussian form is apparent in Fig. \ref{fig7}. We also observe that upon an increase of $t$ the curves get closer to
\begin{align}
P_t(X) = \frac{B}{t_{\rho}^{2/3}} \exp\left(- b \left(\frac{|X|}{t_{\rho}^{2/3}}\right)^{4/3}\right)
\end{align}
for both Model A and Model B. We thus find it absolutely 
plausible that such a form of distribution is also valid for the dynamics of a TP on a populated random Manhattan lattice.

\section{Conclusions}
\label{conc}

To recapitulate, we studied the tracer particle (TP) dynamics in presence of
 two interspersed and
 competing types of disorder - \textit{quenched} random convection flows on a random Manhattan lattice, which prompt the TP to move super-diffusively,  
 and a crowded \textit{dynamical} environment formed by a lattice gas (LG) of hard-core particles, 
  which hinder the TP motion. The random Manhattan lattice is a square lattice decorated with arrows in such a way that directionality of each arrow 
is fixed along each raw (a street) or a column (an avenue) along their entire length, 
but whose orientation randomly  fluctuates from a street to a street and from an avenue to an avenue. 
 
 The hard-core LG particles perform a random motion, constrained by the single-occupancy condition; that being, 
 each lattice site can be occupied by at most a single particle - a LG particle or a TP, or be vacant. 
 We have considered two possible scenarios of the LG particles random motion. 
 In Model A, we supposed that the LG particles are insensitive to the random convection flows and perform symmetric random walks - a simple exclusion process -  among the sites of a two-dimensional square lattice. 
 In this case, the TP moves subject to random convection flows and interacts with a fluid-like quiescent environment, which imposes some frictional force on it. In Model B, we supposed that all the particles - the TP and the LG particles - are identical and follow a local directionality of bonds in a random Manhattan lattice. In this case, the system under study is a kind of a "turbulent" fluid in which all the particles perform a super-diffusive motion. 
 
 We focused on such characteristics of the TP dynamics as its 
 disorder-averaged mean-squared displacement (DA MSD), (and generally, the moments of arbitrary order for a single TP), the distribution of its position at time moment $t$ averaged over disorder, and the time evolution of the kurtosis of this distribution. 
 We have shown that for both Model A and Model B the DA MSD obeys a super-diffusive law $\langle \overline{X_t^2} \rangle \sim m_2(\rho) \, t^{4/3}$, where the functional dependence of the prefactor $m_2(\rho)$ on the mean density $\rho$ 
 of the LG particles depends on the model under study.  
For the case of a single TP (i.e. 
in absence of the LG particles, $\rho = 0$), we provided some analytical arguments explaining such a super-diffusive behaviour. 
 
We showed that the distribution of the TP position has a Gaussian central part and exhibits slower-than-Gaussian tails of the form $\exp(- (|X|/t^{2/3})^{4/3})$ for sufficiently large $X$ and $t$. Such a form was evidenced in case of a single TP through an analysis of a very big statistical sample, and also shown to hold, although 
not in a  completely conclusive way, for the dynamics on a populated random Manhattan lattice.  
 As a consequence of presence of anomalous tails, 
 the kurtosis of the distribution in all the situations under study, was shown to attain a bigger value ($3.5$) than the value ($3$) specific to a Gaussian distribution.
 
 Finally, we addressed the question of sample-to-sample fluctuations in the system under study 
 and performed an analysis of spectral properties of the TP trajectories, which revealed some interesting features.

\section*{Acknowledgments}

The authors acknowledge helpful discussions with E. Agliari, E. Barkai and  A. Gorsky,  and also wish to thank the latter for pointing us on Ref. \cite{ser}. C.M-M and O.V. acknowledge the hospitality of the Interdisciplinary Scientific Center J-V Poncelet (UMI CNRS 2615), Moscow, Russia, where the major part of this work has been done. C.M.M. acknowledges financial support from the Spanish Government Grant No. PGC2018-099944-B-I00 (MCIU/AEI/FEDER, UE).

\end{document}